\documentclass{article}
\usepackage[utf8]{inputenc}
\usepackage{natbib}
\usepackage{amssymb,amsmath,amsthm}
\usepackage{multirow}
\usepackage{graphicx}
\usepackage[margin=1.5in]{geometry}
\usepackage{comment}
\usepackage{color}
\usepackage{algpseudocode}
\usepackage{booktabs}
\usepackage{comment}

\newtheorem{assumption}{Assumption}

\newcommand{\bsx}{{x}}

\newcommand{\bsX}{\boldsymbol{X}}

\newcommand{\e}{\mathbb{E}}
\newcommand{\p}{\mathbb{P}}
\newcommand{\E}{\mathbb{E}}
\newcommand{\var}{\mathrm{var}}

\newcommand{\err}{\varepsilon}

\newcommand{\tran}{\mathsf{T}}

\newcommand{\simiid}{\stackrel{\mathrm{iid}}\sim}

\newcommand{\dnorm}{\mathcal{N}}

\renewcommand{\r}{\mathrm{r}}

\newcommand\indep{\protect\mathpalette{\protect\independenT}{\perp}}
\def\independenT#1#2{\mathrel{\rlap{$#1#2$}\mkern2mu{#1#2}}}

\newcommand{\yit}{Y_i(1)}
\newcommand{\yic}{Y_i(0)}

\newcommand{\bstau}{\boldsymbol \tau}
\newcommand{\hbstau}{\boldsymbol{\hat \tau}}
\newcommand{\taur}{\boldsymbol{\hat \tau_r}}
\newcommand{\tauo}{\boldsymbol{\hat \tau_o}}

\newcommand{\bsxi}{\boldsymbol \xi}
\newcommand{\bsnu}{\boldsymbol \nu}
\newcommand{\bsD}{\boldsymbol d}

\newcommand{\bsSig}{\boldsymbol \Sigma}

\newcommand{\ident}{\boldsymbol I}
\newcommand{\bsdelt}{\boldsymbol \delta}

\newcommand{\bskap}{\boldsymbol \kappa}

\newcommand{\Tr}{\text{tr}}

\newcommand{\diag}{\mathrm{diag}}

\DeclareMathOperator*{\argmin}{arg\,min}

\title{Designing Experiments Toward Shrinkage Estimation}
\author{Evan T. R. Rosenman, Luke Miratrix}

\begin{document}

\maketitle

\abstract{We consider how increasingly available observational data can be used to improve the design of randomized controlled trials (RCTs). We seek to design a prospective RCT, with the intent of using an Empirical Bayes estimator to shrink the causal estimates from our trial toward causal estimates obtained from an observational study. We ask: how might we design the experiment to better complement the observational study in this setting? 

We propose using $\bskap_2$ -- which shrinks each component of the RCT estimator toward its observational counterpart by a factor proportional to its variance -- as our estimator. First, we show that the risk of $\bskap_2$ can be computed efficiently via numerical integration. We then propose algorithms for determining the best allocation of units to strata (the best ``design"). We consider three options: Neyman allocation; a ``naïve” design assuming no unmeasured confounding in the observational study; and a ``defensive" design accounting for the imperfect parameter estimates we would obtain from the observational study with unmeasured confounding. 

We also incorporate results from sensitivity analysis to establish guardrails on the designs, so that our experiment could be reasonably analyzed with and without shrinkage. We demonstrate the superiority of these experimental designs with a simulation study involving causal inference on a rare, binary outcome.
}

\tableofcontents

\section{Introduction}

Recent years have seen increased interest in methods to integrate observational data with experimental data \citep{colnet2020causal}. Such methods have been used to identify average causal effects in target populations \citep{Bareinboim7345, kallus2018removing}, identify heterogeneous treatment effects \citep{peysakhovich2016combining}, and improve precision in causal estimation \citep{gagnon2021precise}

This surge in methodological development is motivated, at least in part, by the increasing proliferation of observational databases. Such repositories provide statisticians with rich new data sources from which to learn. Yet the lurking danger of unmeasured confounding yields rightful trepidation about incorporating these data into estimation procedures \citep{colnet2020causal}. \cite{rosenman2020combining} developed a procedure for shrinking causal estimates from a stratified experiment toward the analogous estimates from an observational study. Shrinkage estimators are attractive in that they allow researchers to use observational data in tandem with experimental data, while protecting the integrity of the randomization of the experiment. They provide a guaranteed reduction in expected loss, relative to using the experimental data alone. 

Separately, \cite{RosenmanOwen+2021+147+171} proposed a method to design more powerful stratified experiments by utilizing information from an observational study to inform decisions about how to allocate a sample across strata and treatment arms. Risk reductions from this method are more modest, owing to the fact that the observational data is used only for design and not for inference. 

Here, we combine the two approaches of design and shrinkage. We seek to design a prospective randomized trial. We optimize the design with two objectives in mind. Our intent is to shrink the causal estimates from our designed trial toward those of an observational study that is already completed. First, we want to design the experiment as a better complement to the observational data, allowing for significant gains in estimation precision. At the same time, we want the experiment to be usable in its own right, and seek to impose guardrails on the design such that the experiment will yield sufficiently precise causal estimates even if we do not elect to include the observational study in the final estimation.



The remainder of this paper proceeds as fellow. In Section \ref{sec:setup}, we define our problem and introduce notation and assumptions. Section \ref{sec:designTowardShrinkage} introduces our choice of shrinkage estimator, $\bskap_2$, and demonstrates how to compute its risk efficiently. This section also discusses three different heuristics under which analysts can design prospective experiments with the intent of leveraging $\bskap_2$ on the final results, while also protecting the utility of the experiment on its own. Section \ref{sec:sims} contains two simulation studies, highlighting the risk improvements that can be attained by designing toward shrinkage. Section \ref{sec:discussion} concludes. 

\section{Set-up}\label{sec:setup}

\subsection{Notation}

We operate in the stratified setting, with fixed strata $k = 1, \dots, K$  defined by covariates. Such strata could be defined by subject matter knowledge, or they could be defined based on application of a modern machine learning method for uncovering heterogeneous treatment effects \citep{hill2011bayesian, wager2018estimation}. 
We begin by setting up the problem in full generality, such that we have access to a pilot dataset that could be either observational or experimental. 

The pilot dataset comprises $n_{\pi}$ total units, indexed by $j$. With each unit, we associate two binary potential outcomes, $Y_j(0)$ and $Y_j(1) \in \{0,1\} $, representing the unit's outcome in the presence and absence of treatment. We also define a treatment variable $W_j \in \{0, 1\}$, representing whether unit $j$ is treated or not; and a vector $\bsX_j \in \mathbb{R}^p$ representing measured covariates. 

The quartets $(\bsX_j, W_j, Y_j(0), Y_j(1))$ are sampled i.i.d. from some population distribution $F_{\pi}$. Denote as $\E_\pi, \var_\pi,$ and $\p_\pi$ the expectation, variance, and probability operators under $F_{\pi}$. Covariate membership is defined by a variable $S_j$ such that $S_j = k \iff \bsX_j \in  \mathcal{X}_k$ for some set of covariate values $\mathcal{X}_k$. 

We design a future blocked experiment of the same treatment. In the experiment, we will recruit a fixed total of $n_{rk}$ units for each stratum $k$, and we will then randomize $n_{rkt}$ of those units to receive the treatment and $n_{rkc} = n_{rk} - n_{rkt}$ to receive the control. We call the list of tuples $\bsD = \{n_{rkt}, n_{rkc}\}_{k = 1}^K$ the ``design." We will assume a sample size constraint such that
\[ \sum_k n_{rkt} + n_{rkc} = n_r \,, \] 
for some fixed number $n_r$. 

Denote as $F_r$ the sampling distribution for the triplets $(X_i, Y_i(0), Y_i(1))$ among units $i$ in the experimental population. We refer to the conditional sampling distribution for units in stratum $k$ as $F_{r \mid S_i = k}$. The experiment can then be understood as drawing $n_{rk}$ units from $F_{r \mid S_i = k}$ and then assigning $W_i$ by choosing a simple random sample of size $n_{rkt}$ from the set of $n_{rk}$ recruited units. We define as $\e_r, \var_\r,$ and $\p_r$ the expectation, variance, and probability operators over both the sampling and treatment randomization in the experiment.

\subsection{Assumptions and Loss Function}

We make the following standard assumption. 

\begin{assumption}[Consistency]\label{ass:consistency}
For each unit $\ell$ in the pilot study or the future experiment, the observed outcome $Y_{\ell} \in \{0, 1\}$ is given by
\[ Y_{\ell} = W_{\ell} Y_{\ell}(1) + (1 - W_{\ell}) Y_{\ell}(0)\,,\] 
that is, there is only one ``version" of the treatment. 
\end{assumption}

A key assumption is that the causal effects across strata are shared between the pilot and future RCT datasets, i.e. 
\begin{assumption}[Common Treatment Effects]\label{assumption:equaltreatmenteffect}
For all strata $k=1,\dots,K$, 
\[  \e_{\pi} \left( Y(1) - Y(0) \mid S = k \right) = \e_r \left( Y(1) - Y(0) \mid S = k \right) = \tau_k\,. \] 
\end{assumption}

Assumption \ref{assumption:equaltreatmenteffect}, sometimes known as the ``transportability condition" \citep{colnet2020causal}, imposes a congruency constraint on the distributions $F_{\pi}$ and $F_r$. This assumption gives us a common target of estimation, which we define as 
\[ \bstau = \left(\tau_1, \dots, \tau_k\right)^{\tran}\,. \] 
We will typically invoke a slightly stronger version of Assumption \ref{assumption:equaltreatmenteffect}, described below. 
\begin{assumption}[Common Moments]\label{assumption:equalMoments}
For all strata $k=1,\dots,K$, and $w \in \{0,1 \}$: 
\[   \e_{\pi} \left( Y(w)  \mid S = k \right) = \e_r \left( Y(w) \mid S = k \right) \hspace{2mm} \text{ and } \hspace{2mm}  \var_{\pi} \left( Y(w)  \mid S = k \right) = \var_r \left( Y(w) \mid S = k \right) \,.\] 
\end{assumption}

We define
\[ \boldsymbol{\hat \tau_{\pi}} = \left( \hat \tau_{\pi 1}, \dots, \hat \tau_{\pi K} \right)^{\tran}, \] 
as the set of stratum causal estimates arising from the pilot study. These estimates can be obtained using a difference-in-means estimator, or a more complex estimator. Though not yet realized, denote as $\taur$ the vector of difference-in-means estimates from the future experiment. 

Our eventual causal estimator, 
\[ \hbstau \equiv \hbstau\left( \taur, \boldsymbol{\hat \tau_{\pi}} \right) =  (\hat \tau_1, \dots, \hat \tau_K)^{\tran},\] 
will be a function of both $\boldsymbol{\hat \tau_{\pi}}$ and $\taur$. Our target of estimation in this manuscript is \emph{not} an average treatment effect over the experimental population. Rather, we seek to obtain good estimates simultaneously for all of the stratum causal effects $\bstau$. Under Assumption \ref{assumption:equaltreatmenteffect}, we can define a loss function under which we evaluate $\hbstau\left( \taur, \boldsymbol{\hat \tau_{\pi}} \right) $. We use the simple, unweighted $L_2$ loss: 
\[ \mathcal{L}(\bstau, \hbstau  ) = \sum_k \left( \tau_k - \hat \tau_k\right)^2\,.  \] 
We seek to optimize the risk of $\hat \bstau$, conditional on $\boldsymbol{\hat \tau_{\pi}}$, which yields the risk expression 
\begin{align*}
\mathcal{R}(\bstau, \hbstau ) &= \E_r \left( \mathcal{L}(\bstau, \hbstau )    \right) \\
&= \sum_k  \E_r \left(\left( \tau_k - \hat \tau_k\right)^2  \right) \,. 
\end{align*}

\subsection{Related Problems}

Under this framing, we can relate this problem to several problems in the causal inference and experimental design. 

First, suppose that the pilot study is an RCT, such that under $F_{\pi}$ we have $W_i \indep Y_i(0), Y_i(1)$. Then, this problem becomes one of adaptive experimental design, in which a trial is conducted in multiple phases and information from the first phase can be used to design later phases. This problem has a rich history in the literature, stretching back to the foundational work of \cite{thompson1933likelihood, thompson1935theory}. While Thompson sampling was originally designed for a more generic problem involving maximizing the expected reward, it can be used to estimate average stratum treatment effects, as discussed in \cite{offer2021adaptive}. Adaptive experimental design is an area of active research, though much recent work has focused on methods that define strata in the second phase, rather than taking strata as fixed \citep{tabord2018stratification, bai2019optimality}. For modern methods that incorporate a fixed stratification scheme, see \cite{hahn2011adaptive} and \cite{chambaz2014targeted}.

Next, suppose that the pilot study is an observational study, but that our ultimate causal estimator $\hbstau$ will simply be $\taur$, the vector of difference-in-means estimators arising from the experiment. This problem is closely related to the survey sampling work of \cite{neyman1992two}. Neyman computed the optimal allocation for a stratified survey -- under a budget constraint -- supposing pilot estimates of variance could be obtained for each stratum. These ideas can easily be extended to the causal inference setting. Here, unbiased pilot stratum variance estimates are obtainable if $W_i \indep Y_i(0), Y_i(1) \mid X_i$ --- that is, unconfoundedness holds in the observational study \citep{Imbens:2015:CIS:2764565}. In our prior work, \cite{RosenmanOwen+2021+147+171}, we demonstrate that efficiency gains are possible even in the presence of unmeasured confounding, as long as we can bound the magnitude of the unmeasured confounding using the sensitivity model of \cite{tan2006distributional}. This is because the observational data can indicate parts of the covariate space where variation is higher or lower -- and hence, where experimenters should over- or undersample.

Lastly, suppose that the pilot study is an observational study, but that the experimental study is already completed, and our goal is to choose an estimator $\hbstau$ to trade off between causal estimates derived from the two data sources.  This is an example of a so-called ``data fusion'' problem \citep{Bareinboim7345} that seeks to merge the observational and experimental data directly. Many methods rely on unconfoundedness in the observational study, including one of our prior papers \citep{rosenman2018propensity} as well as \cite{athey2019surrogate}. Other studies have sought to weaken this condition, and frequently utilize alternative assumptions to proceed with merged estimations.  \cite{kallus2018removing} assumes that the hidden confounding has a parametric structure that can be modeled effectively. \cite{peysakhovich2016combining} propose a method for when the observational data are time series and the bias preserves a unit-level rank ordering. For an excellent overview of the available methods in this area, see \cite{colnet2020causal}.

\subsection{Principles Guiding Estimator Choice and Experimental Design}

In this paper, we will consider the case where the pilot study is an observational study, and the experimental study has yet to be implemented. Hence, we replace the subscript $\pi$ with $o$ (e.g. defining our vector of pilot study causal estimates as $\tauo$ rather than $ \boldsymbol{\hat \tau_{\pi}}$). We will choose an estimator $\hbstau\left( \taur, \tauo \right)$ for combining the observational and experimental data within each stratum. 
Then, we will design our experiment explicitly to minimize the risk of this estimator. In this section, we highlight several characteristics of the estimator and experimental design procedure that we consider ideal.

First, we would like our estimator to be \emph{robust to unmeasured confounding} in the observational study. The assumption of unconfoundedness -- roughly, that all variables affecting both the treatment probability and the outcome have been measured in the observational study -- is fundamentally untestable, and rarely holds in practice \citep{Imbens:2015:CIS:2764565}. 
In this setting, note that our problem is somewhat asymmetric: a simple vector of difference-in-means estimates  from the experimental study will be unbiased, and will be ``good enough" in many cases. Hence, we do not want to incorporate the observational data unless we have strong guarantees that it will reduce statistical risk. Thus, our chosen estimator should not be highly susceptible to bias due to unmeasured confounding in the observational study. 

Second, we would like our procedure to generate experiments that are still valid if they are analyzed alone. We term this feature \emph{detachability}. To motivate this idea, consider an extreme case where we fail to sample any treated or any control units from a particular stratum, under the assumption that the experimental units will simply complement observational units in those strata. Suppose that we later learn of a problem with the observational study that undermines the validity of Assumption \ref{assumption:equaltreatmenteffect} --- e.g., perhaps the data was obtained fraudulently. In this case, we would be out of luck: the experiment has an effective sample size of $0$, so we cannot use it on its own. Some measure of safety could be achieved by imposing design restrictions, such as mandating that each stratum must contain a minimum proportion of treated or control units. But we would like our procedure to naturally regularize toward more equal experimental allocations such that detachability is naturally achieved. 

With these features in mind, we briefly discuss the possibility of using simpler data fusion estimators. In \cite{rosenman2018propensity} we propose a method that is based on simply pooling observational and experimental data within each stratum (termed the ``spiked-in" estimator) and also methods based on convex combinations of $\boldsymbol{\hat \tau_o}$ of $\taur$ (termed the ``weighted average" and ``dynamic weighted average" estimators). These estimators were designed under the assumption that unconfoundedness holds in the observational study. Hence, they do not exhibit the desired robustness property. 
Hence, as discussed in the next section, we turn our attention toward estimators based on Empirical Bayes shrinkage. 

\section{Designing Towards Shrinkage}\label{sec:designTowardShrinkage}

\subsection{Choice of Estimator: $\bskap_{2}$}

In \cite{green1991james} and \cite{green2005improved}, the authors consider how to shrink between an unbiased estimator and a biased estimator. Their goal is to derive Empirical Bayes estimators that guarantee a risk reduction relative to using the unbiased estimator alone. The problem turns out to be quite similar to James-Stein estimation. 

In \cite{rosenman2020combining}, we propose two estimators that are extensions of Green and Strawderman's work. The latter estimator, $\bskap_{2}$ is constructed such that it shrinks each component of the unbiased estimator toward its counterpart in the biased estimator by a factor inversely proportional to its precision. We consider this property desirable in many applied settings. Moreover, this estimator typically outperformed competitor estimators -- including those proposed by Green and Strawderman -- in our applied data analysis. Hence, we proceed in this manuscript with the intention of using $\bskap_{2}$ to shrink between our RCT and observational study estimators. We discuss the use of alternative estimators in the Appendix. 

Let $\taur \in \mathbb{R}^K$ be the unbiased (RCT) estimator and $\tauo \in \mathbb{R}^K$ be the biased (observational study) estimator. Denote as $\boldsymbol \Sigma_r  = \diag(\sigma_{rk}^2) \in \mathbb{R}^{K \times K}$ the diagonal covariance matrix of $\taur$. 
Under mild conditions, we can assume $\taur \sim \dnorm(\bstau, \bsSig_r)$ \citep[see e.g.][]{li2017general}, where $\bstau$ is the vector of true causal estimates. Our estimator is constructed as 
\[ \bskap_{2} = \tauo + \left( \ident_K -  \frac{\Tr(\bsSig_r^2) \bsSig_r}{(\tauo - \taur)^\tran  \bsSig_r^2 (\tauo - \taur)} \right) \left( \taur - \tauo \right) \,.\] 

We can compute, $ \e_r\left( \mathcal{L}(\bstau, \bskap_{2})  \right) $, the risk of $\bskap_{2}$, using results from \cite{strawderman2003minimax}. This yields the expression 

\begin{equation}\label{eq:riskExp}
\begin{aligned}
\mathcal{R}(\bskap_{2}) &= \frac{1}{K} \left( \Tr(\bsSig_r) + \Tr(\bsSig_r^2) \e_r\left(  \frac{4(\taur - \tauo)^{\tran}\bsSig_r^{^4} (\taur - \tauo)}{((\taur - \tauo)^{\tran}\bsSig_r^{^2} (\taur - \tauo))^2} - \frac{ \Tr(\bsSig_r^2) }{(\taur - \tauo)^{\tran}\bsSig_r^{^2} (\taur - \tauo)}  \right) \right)\,,
\end{aligned}
\end{equation}
where the expectation is with respect to $\taur$ only, and $\tauo$ is treated as a constant vector. As shown in \cite{rosenman2020combining}, the estimator is guaranteed to dominate $\taur$ under the squared-error loss $\mathcal{L}(\bstau, \bskap_{2}) = || \bskap_{2} - \bstau ||^2$, as long as the condition 
\begin{equation}\label{improveCondition}
4 \max_k \sigma_{rk}^4 < \sum_k \sigma_{rk}^4\,
\end{equation}
is satisfied. 

We will also frequently consider the positive part analogue of $\bskap_2$,
\[ \bskap_{2+} = \tauo + \left( \ident_K -  \frac{\Tr(\bsSig_r^2) \bsSig_r}{(\tauo - \taur)^\tran  \bsSig_r^2 (\tauo - \taur)} \right)_+ \left( \taur - \tauo \right) \,,\] 
which constrains the shrinkage estimator from ``over-adjusting" by not allowing it to put a negative coefficient on $\tauo$. The risk of $\bskap_{2+}$ is guaranteed to be strictly lower than that of $\bskap_2$. In simulations in \cite{rosenman2020combining}, $\bskap_{2+}$ routinely dominated $\taur$ even when Condition \ref{improveCondition} was not met.


\subsection{Exact Risk Calculation Under Known Parameters}\label{sec:exactRisk}

Our goal is to optimize the experimental design over the risk given in Expression \ref{eq:riskExp}. We suppose, for a moment, that the quantities $\bsSig_r$ and $\bsxi = \e_r(\taur -  \tauo)$ are known. Expression \ref{eq:riskExp} can then be rewritten as 
\[ \mathcal{R}(\bskap_{2}) = \frac{1}{K} \left( \Tr(\bsSig_r) + \Tr(\bsSig_r^2) \e_r\left(  \frac{4\cdot \bsnu^{\tran}\bsSig_r^{^5} \bsnu}{(\bsnu^{\tran}\bsSig_r^{^3} \bsnu)^2} - \frac{ \Tr(\bsSig_r^2) }{\bsnu^{\tran}\bsSig_r^{^3} \bsnu}  \right) \right)\,, \] 
where $\bsnu \sim \dnorm(\bsSig_r^{-1/2}\bsxi, \bsSig_r)$. 

The expectation is a difference of first moments of ratios of quadratic forms in normal random variables. Exact expressions for these integrals can be found in \cite{bao2013moments}. In particular,

\begin{equation}\label{eq:integrals}
\begin{aligned}
\e_r\left(  \frac{ \bsnu^{\tran}\bsSig_r^{^5} \bsnu}{(\bsnu^{\tran}\bsSig_r^{^3} \bsnu)^2} \right) &= \int_0^{\infty} \det ( \boldsymbol I_K + 2 t \bsSig_r^3 )^{-1/2} \cdot \exp\left( \frac{1}{2} \left(  \bsxi^\tran ( \boldsymbol I_K + 2 t \bsSig_r^3 )^{-1} \bsxi - \bsxi^{\tran} \bsxi \right) \right) \cdot \\
& \hspace{11mm}\left( \Tr (\boldsymbol R) + (\boldsymbol L \bsSig_r^{-1/2} \bsxi)^\tran \boldsymbol R (\boldsymbol L \bsSig_r^{-1/2} \bsxi) \right) t dt \\
\e_r\left(  \frac{1}{(\bsnu^{\tran}\bsSig_r^{^3} \bsnu)} \right) &= \int_0^{\infty} \det ( \boldsymbol I_K + 2 t \bsSig_r^3 )^{-1/2} \cdot \exp\left( \frac{1}{2} \left(  \bsxi^\tran ( \boldsymbol I_K + 2 t \bsSig_r^3 )^{-1} \bsxi - \bsxi^{\tran} \bsxi \right) \right) dt 
\end{aligned}
\end{equation}
where 
\begin{align*}
\boldsymbol  L &= (\boldsymbol I_K + 2 t \bsSig_r^3)^{-1/2}, \hspace{5mm} \text{ and } \\
\boldsymbol R &=  \boldsymbol  L ^{\tran} \bsSig_r^5 \boldsymbol{L}. 
\end{align*}

The integrals in Expression \ref{eq:integrals} can be computed via numerical integration, yielding an efficient evaluation of the risk for each possible choice of the parameter values.

\subsection{Design Options}\label{sec:designOptions}
In this section, we consider several heuristics for designing the experiment in the absence of perfect knowledge of $\bsSig_r$ and $\bsxi$. 

\subsubsection{Neyman Allocation}

Because the observational data is not randomized, we will need to conduct some form of statistical adjustment to remove confounding bias in $\tauo$. In this manuscript, we will use stabilized inverse probability of treatment weighting (SIPW) as our adjustment method. Briefly, this involves estimating the propensity score -- the probability of treatment in the observational study -- as a function of the observed covariates, and scaling the observed outcomes by the inverse of each unit's estimated propensity score. For more details, see \cite{Imbens:2015:CIS:2764565}. 

Once the adjustment is made, a simple approach to the experimental design problem is to assume that there is no residual unmeasured confounding in the observational study. Importantly, we can make this assumption for the purposes of \emph{design} only.
The unconfoundedness assumption need not be strictly true to ensure good performance of $\bskap_{2}$ once our experiment is completed. The shrinkage properties of $\bskap_{2}$ ensure that its risk will be lower than $\taur$ as long as Condition \ref{improveCondition} is met, irrespective of the presence of residual confounding in the observational study. Hence, we retain the implicit guarantee against a risk increase when it comes to \emph{inference}, even if this assumption turns out to be incorrect. 


If we are willing to suppose that Assumption \ref{assumption:equalMoments} holds, as well as unconfoundedness, we can use the observational study to unbiasedly estimate the mean and variance of the potential outcomes in each stratum. Denote the estimates obtained from the observational data as  
\begin{align*}
\hat \sigma_{kt}^2 &= \widehat \var \left( Y(1) \mid S = k \right) \hspace{5mm} \text{ and } \hspace{5mm} \hat \sigma_{kc}^2 = \widehat \var \left( Y(0) \mid S = k \right) \,. 
\end{align*}

The simplest design heuristic is to then use a Neyman allocation \citep{splawa1990application} without a cost constraint, e.g. 
\begin{align*}
n_{rkt} = \frac{n_r \cdot\hat \sigma_{kt}^2}{\sum_k\hat \sigma_{kt}^2 +  \hat \sigma_{kc}^2}\ \hspace{5mm} \text{ and } \hspace{5mm} n_{rkc} &= \frac{n_r \cdot \hat \sigma_{kc}^2}{\sum_k\hat \sigma_{kt}^2 +\hat \sigma_{kc}^2}\,.
\end{align*}
Such a design would be optimal if the risk of our estimator were only $\Tr(\bsSig_r)/K$, the first term in Expression \ref{eq:riskExp}. Though the design does not directly optimize over the shrinkage portion of the risk, it serves as a reasonable starting point for the purposes of design. As we will see in Section \ref{sec:sims}, it also typically yields good performance for $\bskap_{2}$ in simulations. 

\subsubsection{Heuristic Optimization Assuming $\bsxi \boldsymbol{= 0}$}

Per the discussion in Section \ref{sec:exactRisk}, we can compute the risk exactly if both $\bsSig_r$ and $\bsxi = \e_r(\taur) - \tauo$ are known. Under Assumption \ref{assumption:equalMoments} and unconfoundedness, $\bsSig_r$ can be estimated unbiasedly for any choice of $\bsD = \{(n_{rkt}, n_{rkc})\}_{k}$. However, $\bsxi$ may be nonzero even if unconfoundedness holds, because we consider $\tauo$ to be a fixed draw from the observational distribution, rather than a random variable. 

In this section, we make the additional assumption that $\bsxi = 0$. Then, we have all the necessary parameter estimates to optimize $ \mathcal{R}(\bskap_{2}) $ over the choice of $\bsD = \{(n_{rkt}, n_{rkc})\}_{k}$. This problem is directly encoded in Optimization Problem \ref{optProb:riskExpNoConfound}:

\begin{equation}\label{optProb:riskExpNoConfound}
\begin{aligned}
\text{ minimize} & \hspace{5mm} \mathcal{R}(\bskap_{2}) \\
\text{ subject to} & \hspace{5mm} \sigma_{rk}^2 = \frac{\widehat \var_r \left( Y(1) \mid S = k \right)}{n_{rkt}} + \frac{\hat \sigma_{kc}^2}{n_{rkc}}, \hspace{2mm} k = 1, \dots, K, \\
& \hspace{5mm} 0 < n_{rkt}, n_{rkc}, , \hspace{2mm} k = 1, \dots, K, \\
& \hspace{5mm}n_r = \sum_k n_{rkt} + n_{rkc}\,,
\end{aligned}
\end{equation} 

Unfortunately, $ \mathcal{R}(\bskap_{2})$ is not convex in $d$. However, Optimization Problem \ref{optProb:riskExpNoConfound} can be approximately solved using a greedy algorithm. Define 
\[ \bsD_j = \{ (n_{rkt}^{(j)}, n_{rkc}^{(j)}\}_k \] 
as the allocation of RCT units to strata and treatment level at iteration $j$ of the algorithm. Next, define 
\[ D_j = \{ \bsD \mid \text{ $\bsD$ swaps exactly one unit across strata or treatment level from } \bsD_j \}\,. \] 
Because there are $K$ strata and two treatment levels, the ``swap set" $D_j$ will typically contain $2K \times (2K - 1)$ possible allocations. This cardinality can be reduced by imposing additional constraints, as will be discussed in Section \ref{sec:guardrails}. 

Define $R(\bsD, \boldsymbol V, \bsxi)$ as the value of $\mathcal{R}(\bskap_{2}) $ evaluated under the design $\bsD$ with estimated stratum potential outcome variances $\boldsymbol V$ and error vectors $\bsxi$. We will evaluate $R(\bsD, \boldsymbol V, \bsxi)$ under estimated variances $\boldsymbol V^{\star} = \left(\hat \sigma_{kt}^2, \hat \sigma_{kc}^2 \right)_{k = 1}^K$ and $\bsxi = 0$. 

Now, we can approximately solve Optimization Problem \ref{optProb:riskExpNoConfound} using Algorithm \ref{greedyAlgorithm}: 

\begin{equation}\label{greedyAlgorithm}
\begin{aligned}
& \texttt{Start with design $\bsD_0 = \{ (n_{rkt}^{(0)}, n_{rkc}^{(0)}\}_k$.} \hspace{55mm} \\
& \texttt{For iteration $j = 1, 2, \dots$:}\\
& \hspace{5mm} \texttt{For each design $\bsD$ in $D_{j-1}$:}\\
& \hspace{10mm} \texttt{Compute $R(\bsD, V^{\star}, 0)$. } \\
& \hspace{5mm} \texttt{Set $\bsD_j = \argmin_{\bsD \in D_{j - 1}} R(\bsD)$} \\
& \hspace{5mm} \texttt{If $R(\bsD_j, V^{\star}, 0) >=  R(\bsD_{j - 1}, V^{\star}, 0)$}\\
& \hspace{10mm} \texttt{Return $\bsD_{j - 1}.$}
\end{aligned}
\end{equation}

In words, Algorithm \ref{greedyAlgorithm} will continue to swap units between strata and treatment levels until no swap will further reduce the estimated risk of the shrinkage estimator. The algorithm naturally enforces the sample size constraint and ensures that the returned values will be integers. 

Because the objective is non-convex, it is plausible that Algorithm \ref{greedyAlgorithm} could get stuck at a local optimum. Practically, we recommend running the algorithm at a few different starting points (e.g. an equally allocated design, the Neyman allocation, and several randomly chosen designs), and choose the design which achieves the minimum final value of the risk. While this approach is not guaranteed to find an optimum, it will nonetheless find a point with a reduced value of the objective function. If the relevant assumptions hold, this point will also be guaranteed to improve efficiency relative to a Neyman allocation. 

\subsubsection{Heuristic Optimization Assuming Worst-Case Error Under $\boldsymbol \Gamma$-Level Unmeasured Confounding}\label{subsec:tan}

The assumption that $\bsxi = 0$ is fundamentally optimistic: it is unlikely to hold even in the absence of unmeasured confounding. If there are unmeasured confounders, it may be far from the truth. We can take a more defensive approach by imposing a sensitivity model on the observational study, and optimizing under the worst-case choice of $\bsxi$. 

As in \cite{RosenmanOwen+2021+147+171}, we constrain the magnitude of the unmeasured confounding by imposing the marginal sensitivity model of \cite{tan2006distributional}. Under this model, a key odds ratio -- between the treatment probability conditional on the potential outcomes and covariates and the treatment probability conditional on covariates only -- is bounded between $1/\Gamma$ and $\Gamma$, for a user-chosen parameter $\Gamma \geq 1$.  The Tan model can be seen as extending the popular Rosenbaum sensitivity model \citep{rosenbaum1987sensitivity} to the setting of inverse probability weighting. In practice, $\Gamma$ can be chosen by computing the treatments odds ratio with and without conditioning on each of the \emph{measured} covariates, and choosing the maximum.

Under Assumption \ref{assumption:equalMoments}, we have  
\begin{align*}
\mu_{kt} &\equiv \e_o(Y(1) \mid S= k) = \e_R(Y(1) \mid S= k), \hspace{5mm} \text{ and } \\
\mu_{kc} &\equiv \e_o(Y(0) \mid S= k) = \e_R(Y(0) \mid S= k),
\end{align*}
for $k = 1, \dots, K$. For any choice of Type I error bound $\alpha \in (0, 1)$, we can use the method of \cite{zhao2019sensitivity} to obtain intervals $\left(\ell_{kt}^{(\Gamma, \alpha)}, u_{kt}^{(\Gamma, \alpha)}\right)$ and $\left(\ell_{kc}^{(\Gamma, \alpha)}, u_{kc}^{(\Gamma, \alpha)} \right)$ such that parameters $\mu_{kt}$ and $\mu_{kc}$ reside within the intervals with at least $1 - \alpha$ probability as long as the true confounding structure lies within the sensitivity model parameterized by $\Gamma$. The method relies on convex optimization and the bootstrap in order to generate valid confidence sets.

Because our outcome is binary, we can then use the confidence sets on $\mu_{kt}$ and $\mu_{kc}$ to obtain valid confidence sets for the stratum potential outcome variances $\sigma_{kt}^2$ and $\sigma_{kc}^2$ via the relations 
\[ \sigma_{kt}^2 = \mu_{kt} \cdot ( 1 - \mu_{kt} ) \hspace{5mm} \text{ and } \hspace{5mm}   \sigma_{kc}^2 = \mu_{kc} \cdot ( 1 - \mu_{kc} )\,. \] 
For a full justification, see  \cite{RosenmanOwen+2021+147+171}. 

Putting these ideas together: under our calibrated choice of $\Gamma$ and a reasonable choice of $\alpha$, we set 
\begin{align*}
\xi_k &= \max\left( \left| u_{kt}^{(\Gamma, \alpha)} - \ell_{kc}^{(\Gamma, \alpha)} - \hat \tau_{ok} \right|, \left| \ell_{kt}^{(\Gamma, \alpha)} - u_{kc}^{(\Gamma, \alpha)} - \hat \tau_{ok} \right| \right), 
\end{align*}
the worst-case value of the error under our sensitivity model. We collect these quantities into a vector $\bsxi'$. Next, we collect the corresponding values of the variances, e.g. 
\[ \hat \sigma_{kt}^2 = \left\{ \begin{array}{cc} u_{kt}^{(\Gamma, \alpha)} \cdot (1 - u_{kt}^{(\Gamma, \alpha)}) & \text{if }  \left| u_{kt}^{(\Gamma, \alpha)} - \ell_{kc}^{(\Gamma, \alpha)} - \hat \tau_{ok} \right| > \left| \ell_{kt}^{(\Gamma, \alpha)} - u_{kc}^{(\Gamma, \alpha)} - \hat \tau_{ok} \right|  \\
\ell_{kt}^{(\Gamma, \alpha)} \cdot (1 - \ell_{kt}^{(\Gamma, \alpha)})  &\text{otherwise} \end{array} \right. \] 
and 
\[ \hat \sigma_{kc}^2 = \left\{ \begin{array}{cc} \ell_{kc}^{(\Gamma, \alpha)} \cdot (1 - \ell_{kc}^{(\Gamma, \alpha)}) & \text{if }  \left| u_{kt}^{(\Gamma, \alpha)} - \ell_{kc}^{(\Gamma, \alpha)} - \hat \tau_{ok} \right| > \left| \ell_{kt}^{(\Gamma, \alpha)} - u_{kc}^{(\Gamma, \alpha)} - \hat \tau_{ok} \right|  \\
u_{kc}^{(\Gamma, \alpha)} \cdot (1 - u_{kc}^{(\Gamma, \alpha)})  &\text{otherwise} \end{array} \right. \] 
into a matrix $\boldsymbol V'$.

Finally, we can evaluate our function $R(\bsD, \boldsymbol V', \bsxi')$ to obtain the risk of $\bskap_{2}$ for any experimental design $\bsD$ under these parameters. The procedure is henceforth exactly analogous to the one used in the prior section: we run Algorithm \ref{greedyAlgorithm}, substituting $R(\bsD, \boldsymbol V', \bsxi')$ for $R(\bsD, \boldsymbol V, 0)$, and obtain the design that yields the lowest value of the risk. 


\subsection{Imposing Guardrails on Designs}\label{sec:guardrails}

Algorithm \ref{greedyAlgorithm} can easily incorporate further restrictions on the set of possible designs. The simplest method is to set the risk computation for any invalid design equal to infinity. This will force the algorithm not to choose that design on the next iteration. 

Practically speaking, we suggest incorporating three constraints on the set of possible designs. First, we suggest imposing a \emph{minimum sample size constraint} such that the allocations to any stratum and treatment group cannot be lower than some value $\text{SS}_{\text{min}}$. This constraint serves two purposes. First, the risk expression in Equation \ref{eq:riskExp} is dependent on the normality of $\taur$. In simulations, we find that the risk expression is quite robust to modest deviations from normality, but sample sizes could plausibly become so small that a Central Limit Theorem need not hold even approximately. Second, this constraint naturally helps improve detachability, since it prevents the variance of any entry of $\taur$ from growing too large. 

Second, we suggest imposing an explicit \emph{detachability} constraint. Inherent in this constraint is the idea that an analyst wants to ensure that the experiment can be analyzed on its own in case stakeholders ultimately decide not to use the observational data at all. One way to impose this constraint is to assume the observational study point estimates of the strata potential outcome variances are correct, and to choose some baseline design $\boldsymbol{\tilde d} =  \{\tilde n_{rkt}, \tilde n_{rkc}\}_k$, e.g. equal allocation or Neyman allocation. We then invalidate any design $\boldsymbol{d'} = \{n_{rkt}', n_{rkc}'\}_k$ for which 
\[ \sum_k  \frac{\hat \sigma_{kt}^2}{n_{rkt}'} + \frac{\hat \sigma_{kc}^2}{n_{rkc}'} \geq \delta_d \sum_k  \frac{\hat \sigma_{kt}^2}{\tilde n_{rkt}} + \frac{\hat \sigma_{kc}^2}{\tilde n_{rkc}}\,, \] 
where $\delta_d \geq 1$ is some user-chosen tolerance parameter. In words, this approach invalidates any design $\boldsymbol{d'}$ for which the estimated risk of $\taur$ under design $\boldsymbol{d'}$ is larger than the estimated risk under the default design by a factor greater than $\delta_d$. 

A more robust version of the detachability constraint can be imposed if the analyst assumes the Tan sensitivity model, as discussed in Section \ref{subsec:tan}. Under a given choice of $\Gamma$ and $\alpha$, we can obtain bounds on the potential outcome means in each stratum. We reject any design such that 
\begin{equation}\label{eq:robustDetach}
 \max_{\mu_{kc} \in \left(\ell_{kc}^{(\Gamma, \alpha)} u_{kc}^{(\Gamma, \alpha)} \right), \atop \mu_{kt} \in \left( \ell_{kt}^{(\Gamma, \alpha)}, u_{kt}^{(\Gamma, \alpha) }\right)} \frac{ \sum_k  \frac{\mu_{kt} ( 1 - \mu_{kt}) }{n_{rkt}'} + \frac{\mu_{kt} ( 1 - \mu_{kt})}{\tilde n_{rkc}} }{ \sum_k  \frac{\mu_{kt} ( 1 - \mu_{kt})}{n_{rkt}'} + \frac{\mu_{kc} ( 1 - \mu_{kc})}{\tilde n_{rkc}} }  \geq \delta_d\,.
\end{equation}
In words, this means that we are rejecting  any design $\boldsymbol{d'}$ such that the risk of $\taur$ under design $d'$ is larger than the estimated risk under the default design by a factor greater than $\delta_d$, for \emph{any} configuration of the potential outcome means consistent with our sensitivity bounds. Practically the left-hand-side of Inequality \ref{eq:robustDetach} can be reduced to a quadratic fractional programming problem and solved via Dinkelbach's method \citep{dinkelbach1967nonlinear}. 

Third, we suggest imposing a \emph{risk reduction} constraint. This means that we enforce Condition \ref{improveCondition} explicitly by not allowing the optimization algorithm to choose allocations under which it fails to hold. Like the detachability constraint, we can impose either a point estimate version or a robust version. The point estimate version invalidates any design  design $\boldsymbol{d'} = \{n_{rkt}', n_{rkc}'\}_k$ for which 
\begin{align*}
4 \max_k & \left( \frac{\hat \sigma_{kt}^2}{n_{rkt}'} + \frac{\hat \sigma_{kc}^2}{n_{rkc}'} \right)^2 >  
 \sum_k \left(  \frac{\hat \sigma_{kt}^2}{n_{rkt}'} + \frac{\hat \sigma_{kc}^2}{n_{rkc}'} \right)^2. 
\end{align*}
The robust version incorporates the parameter bounds from the Tan sensitivity model. We reject any design such that 
\begin{align*}
4  \max_k & \min_{\mu_{kc} \in \left(\ell_{kc}^{(\Gamma, \alpha)} u_{kc}^{(\Gamma, \alpha)} \right), \atop \mu_{kt} \in \left( \ell_{kt}^{(\Gamma, \alpha)}, u_{kt}^{(\Gamma, \alpha) }\right)}   \left( \frac{\mu_{kt} ( 1- \mu_{kt})}{n_{rkt}'} + \frac{\mu_{kc} (1 - \mu_{kc})}{n_{rkc}'} \right)^2 >  \\
&  \max_{\mu_{kc} \in \left(\ell_{kc}^{(\Gamma, \alpha)} u_{kc}^{(\Gamma, \alpha)} \right), \atop \mu_{kt} \in \left( \ell_{kt}^{(\Gamma, \alpha)}, u_{kt}^{(\Gamma, \alpha) }\right)} \sum_k \left( \frac{\mu_{kt} ( 1- \mu_{kt})}{n_{rkt}'} + \frac{\mu_{kc} (1 - \mu_{kc})}{n_{rkc}'} \right)^2. 
\end{align*}

\section{Simulation Study}\label{sec:sims}

\subsection{Simulation Set-Up}

We pattern the simulations on those in \cite{rosenman2018propensity}, considering a situation with a relatively rare, binary outcome and a modest effect size. 

 We first generate an observational super-population of $1 \times 10^6$ units, and an experimental super-population of the same size. We suppose that the completed observational study comprises $20,000$ units, sampled a single time from the corresponding super-population. The prospective RCT comprises $1,000$ units, which will be sampled repeatedly from the experimental super-population. 

We define $j \in \mathcal{O}$ as the indexing variable for the observational super-population and $i \in \mathcal{E}$ as the indexing variable for the experimental super-population. We use $\ell$ as an index over both populations. For each unit $\ell \in \mathcal{O} \cup \mathcal{E}$, we suppose there is a covariate vector $\bsX_\ell \in \mathbb{R}^5$ where $\bsX_\ell \simiid \mathcal{N}\left(0, \Sigma\right)$ for $\Sigma$ such that each covariate has unit variance and roughly a quarter of the covariances are $+0.1$, roughly a quarter are $-0.1$, and the remainder are $0$. Such a covariance structure is roughly consistent with the applied data analysis from the Women's Health Initiative, as used in \cite{rosenman2018propensity}. 

The untreated potential outcomes $Y_\ell(0)$ are sampled as independent Bernoulli random variables with 
\[\text{Pr}(Y_\ell(0) = 1 \mid \bsx_\ell) = \frac{1}{1 + e^{-\alpha -\beta^\tran\bsx_\ell + \err_\ell}} ,\quad\text{for $\beta = (1,1,1,1,1)^\tran$}\] 
where $\err_\ell$ are generated as IID $\dnorm(0,1)$ random variables and $\alpha$ is chosen such that the average incidence rate is 10\%. The treatment variables  in the observational study are independent Bernoulli random variables with
\[\text{Pr}(W_j = 1 \mid \bsx_j) = \frac{1}{1 + e^{-\gamma^\tran\bsx_j}},\quad\text{for $\gamma = (\sqrt{2}, \sqrt{2}, \sqrt{2}, 0, 0)^\tran$.}\] 
The nonzero inner product between $\beta$ and $\gamma$ induces strong selection bias in the observational study, with treated units likelier to have larger untreated potential outcomes. 

For both datasets, we suppose the data is split into twelve strata based on the first and second covariates. The strata boundaries are defined by the 25th and 50th quantiles of the first covariate and by the quartiles of the second covariate. We seek to obtain estimates of the average causal effect in each of the resultant strata. 

In both datasets, we assign individual treatment effects to match three different structures. We number the strata from $k = 1, \dots, 12$ such that stratum $1$ corresponds to the lowest stratum on both covariates, stratum 2 corresponds to the lowest stratum on the first covariate and the second quartile of the second covariate, stratum $3$ corresponds to the lowest stratum on the first covariate and the third quartile of the second covariate, etc. The values of $\tau_k$, the stratum average treatment effects, in the constant, linear, and quadratic strata-level treatment effect models are
\begin{align}\label{eq:treatments}
\tau_k  = T, \quad \tau_k = -T\times \frac{k}{K},\quad\text{and}\quad
\tau_k  = T\times \left(\frac{k}{K}\right)^2
\end{align}
respectively. In each case we choose the scale $T>0$ so that Cohen's $d$ \citep{cohen1988statistical} in the observational study precisely equals $0.5$, which Cohen calls a medium effect size. 

Denote as $n_k$ the total number of super-population units that fall into a given stratum. Given the $\tau_k$, the treated potential outcomes are assigned by randomly selecting $\tau_k \times n_k$ units for which $Y_{\ell}(0) = 0$, and setting $Y_{\ell}(1) = 1$ for those units. For all of the remaining units, $Y_{\ell}(0) = Y_{\ell}(1)$. Because we use the same process across both super-populations, we enforce Assumption \ref{assumption:equalMoments}, such that the stratum-specific causal effects (as well as potential outcome means and variances) are shared under the observational and RCT data distributions.


The observational data is sampled a single time. Next, leveraging the observational sample, we compute the allocations of units to strata in the RCT (the RCT ``design") using each of the methods discussed in the prior section. For each possible design, we sample the RCT units from the super-population $5,000$ times. For each iteration, for each stratum $k$, we assume treatment is assigned via a simple random sample of $n_{rkt}$ units out of the $n_{rkt} + n_{rkc}$ units recruited for the stratum. Once the units are drawn and treatments are assigned, we compute the estimators $\taur$, $\bskap_2$, and $\bskap_{2+}$. We compute the $L_2$ distance between each estimate and the true treatment effects $\tau$, and take the average over all $5,000$ simulations.

\subsection{Ideal Case: No Unmeasured Confounding}

We begin with the simplest case: we suppose all of the covariates are measured in the observational study, so there is no residual unmeasured confounding. This is an idealized case, in which all of the selection bias in the observational study can be removed with a statistical adjustment. We fit a propensity score to the observational study data and use stabilized inverse probability of treatment weighting (SIPW) to compute the observational causal estimates.

In Table \ref{tab:idealized}, we show the average $L_2$ errors over $5,000$ simulations. We consider the equal allocation and Neyman allocation designs, as well as the ``naive" design assuming $\bsxi = 0$. We also consider the defensive approach discussed in Section \ref{subsec:tan}, and compute the optimal design under errors computed with the Tan sensitivity model parameter set to 1.0, 1.1, 1.2, and 1.5. Lastly, we consider an ``oracle" design, in which we run Algorithm \ref{greedyAlgorithm} but provide it with the true values of the potential outcome variances and the error in the observational study. Results are given for the three different treatment effect models: constant (c), linear ($\ell$), and quadratic (q). Risk estimates are expressed as percentages of the risk of $\taur$ when using an equal allocation for the given treatment effect model. 

\begin{table}[h]
\centering
\begin{tabular}{ll|rrr|rrrr|r}
\toprule
 & & & & &  \multicolumn{4}{c|}{\textbf{Max Bias, }$\boldsymbol \Gamma$ \textbf{Value} } &  \\ 
\textbf{Est.}  & \textbf{Trt.}    & \textbf{Equal} & \textbf{Neyman} & \textbf{Naïve} & \textbf{1.0} & \textbf{1.1} & \textbf{1.2} & \textbf{1.5} & \textbf{Oracle} \\ \midrule
$\taur$   & \multirow{3}{*}{c} & 100\% & \underline{87\%} & 91\% & 100\% & 96\% & 94\% & 94\% & 96\% \\
$\bskap_2$      &  & 82\% & 48\% & \underline{44\%} & 52\% & 48\% & 47\% & 50\% & 42\% \\
$\bskap_{2+}$ & &38\% & 28\% & \underline{26\%} & 26\% & 26\% & 26\% & 28\% & 23\% \\ \midrule
$\taur$   &  \multirow{3}{*}{$\ell$} & 100\% & \underline{89\%} & 92\% & 95\% & 94\% & 95\% & 97\% & 104\% \\
$\bskap_2$      &  & 93\% & 66\% & 58\% & 58\% & \underline{57\%} & 60\% & 64\% & 50\% \\
$\bskap_{2+}$ & & 59\% & 51\% & 45\% & \underline{43\%} & 45\% & 47\% & 49\% & 33\% \\ \midrule				
$\taur$   & \multirow{3}{*}{q} & 100\% & \underline{86\%} & 91\% & 95\% & 98\% & 94\% & 92\% & 91\% \\
$\bskap_{2+}$  & &81\% & 47\% & \underline{45\%} & 52\% & 52\% & 50\% & 48\% & 41\% \\
$\bskap_{2+}$ & & 37\% & 29\% & \underline{27\%} & 28\% & 28\% & 30\% & 29\% & 25\% \\ \bottomrule
 \end{tabular} 
\caption{\label{tab:idealized} Risk over $5,000$ iterations of $\taur, \bskap_2$, and $\bskap_{2+}$ under various experimental designs, in the case of no unmeasured confounding in the observational study. Risks are expressed as a percentage of the risk of $\taur$ using an equally allocated experiment, for each of the three treatment effect models. The minimum non-oracle risk in each row is denoted with an underline.}
\end{table}

Across the three treatment effect models, we see that results are relatively consistent in the ordering of the estimators. The Neyman allocation always performs best when using $\taur$ alone, realizing an 8$-$14\% error reduction relative to the equal allocation. The other allocations typically achieve more modest error reductions when using $\taur$. 

If we stick with the Neyman allocation but switch to using either the shrinkage estimator $\bskap_2$ or its positive part analogue $\bskap_{2+}$, we can realize massive additional error reductions on the order of 45 to 60\%, depending on the treatment effect model. However, the na\"ive allocation is typically even better, allowing us to realize improvements of a few additional percentage points in error. The na\"ive allocation yields the best performance with $\bskap_2$ and $\bskap_{2+}$ in the linear and quadratic treatment effect models, with the robust allocations typically doing nearly as well. In the linear treatment effect model, we actually do best using a $\Gamma = 1.1$ robust design for $\bskap_2$ and a $\Gamma = 1.0$ robust design for $\bskap_{2+}$, with the na\"ive allocation performing nearly as well. 


\subsection{Practical Case: Unmeasured Confounding} 

The assumption of unconfoundedness is not defensible in many practical settings, in which a relatively sparse set of covariates are measured in the observational study. Moreover, the assumption is not testable. Hence, we run a second set of simulations in which we induce confounding by assuming that the third entry in $\bsx_i$ is not measured. This third covariate affects both treatment probabilities and outcomes, so bias in the observational study cannot be fully corrected with a propensity score adjustment.

Results are given in Table \ref{tab:practical}. Results differ slightly from the simulations assuming no unmeasured confounding. When using $\bskap_2$ and $\bskap_{2+}$, the best non-oracle designs are the robust designs assuming $\Gamma$-level unmeasured confounding for $\Gamma \in \{1.0, 1.1, 1.2\}$. The design under $\Gamma = 1$ does best in the constant and linear case, while the design under $\Gamma = 1.2$ does best in the quadratic case. The na\"ive allocation still performs quite well, with risk typically only a few percentage points worse than the best performer. 


\begin{table}[h]
\centering
\begin{tabular}{ll|rrr|rrrr|r}
\toprule
 & & & & &  \multicolumn{4}{c|}{\textbf{Max Bias, }$\boldsymbol \Gamma$ \textbf{Value} } &  \\ 
\textbf{Est.}  & \textbf{Trt.}    & \textbf{Equal} & \textbf{Neyman} & \textbf{Naïve} & \textbf{1.0} & \textbf{1.1} & \textbf{1.2} & \textbf{1.5} & \textbf{Oracle} \\ \midrule
$\taur$   & \multirow{3}{*}{c}& 100\% & \underline{90\%} & 90\% & 90\% & 92\% & 93\% & 95\% & 102\% \\
$\bskap_2$      &  &102\% & 81\% & 74\% & \underline{72\%} & 72\% & 72\% & 77\% & 69\% \\
$\bskap_2$      &  &96\% & 80\% & 74\% & \underline{71\%} & 72\% & 72\% & 76\% & 67\% \\ \midrule
$\taur$   & \multirow{3}{*}{$\ell$} & 100\% & 93\% & \underline{93\%} & 94\% & 95\% & 96\% & 96\% & 104\% \\
$\bskap_2$      & &102\% & 85\% & 77\% & \underline{75\%} & 76\% & 77\% & 79\% & 73\% \\
$\bskap_{2+}$ & & 98\% & 84\% & 77\% & \underline{75\%} & 76\% & 76\% & 79\% & 71\% \\ \midrule
$\taur$   & \multirow{3}{*}{q} &100\% & \underline{89\%} & 90\% & 93\% & 92\% & 91\% & 96\% & 96\% \\
$\bskap_{2+}$ & &101\% & 74\% & 69\% & 68\% & 68\% & \underline{67\%} & 73\% & 66\% \\ 
$\bskap_{2+}$ & &88\% & 72\% & 67\% & 66\% & 66\% & \underline{65\%} & 71\% & 63\% \\ \bottomrule
 \end{tabular} 
\caption{\label{tab:practical} Risk over $5,000$ iterations of $\taur, \bskap_2$, and $\bskap_{2+}$ under various experimental designs, in the case of unmeasured confounding in the observational study via failure to measure the third covariate. Risks are expressed as a percentage of the risk of $\taur$ using an equally allocated experiment, for each of the three treatment effect models. The minimum non-oracle risk in each row is denoted with an underline.}
\end{table} 

These results point to a few practical guidelines for designing toward shrinkage. First, we find that the na\"ive allocation is quite robust, even when $\bsxi$ is far from zero. This is evident from the fact that the na\"ive allocation always yields significant performance gains over the equal and Neyman allocations when using a shrinkage estimator, even when unmeasured confounding is present.

Second, in the presence of unmeasured confounding, one may achieve modest further improvements from enforcing a robust design at a relatively low value of $\Gamma$. In this example, we have \emph{not} constructed the confounding such that it matches the form of our sensitivity model. Exclusion of the third covariate induces enormous discrepancies in the treatment odds between the true and estimated propensity scores in a small proportion of individuals: for about 0.2\% of the super-population units, the difference exceeds a multiplicative factor of 100. For most of the population, however, the true and estimated treatment odds differ by a much smaller factor. The Tan model imposes a worst-case bound on the deviation between the true and estimated odds of treatment, so no value of $\Gamma$ between 1.0 and 2.0 is large enough to account for our most extreme deviations. 

Nonetheless, the magnitude of the worst-case error under the Tan model correlates reasonably well with the true values of $\bsxi$  when choosing $\Gamma = 1.0, 1.1,$ or $1.2$, offering one explanation for the strong performance of the allocations designed under these schemes. In a more general sense, the robust allocations under $\Gamma$ serve as a form of regularization, bringing the design closer to an equal allocation to hedge against the possibility that $\tauo$ is far off from $\bstau$. In doing so, the robust designs may improve empirical performance even when the Tan model does not accurately characterize the form of unmeasured confounding. 


\section{Discussion}\label{sec:discussion} 

We have considered the problem of designing a stratified experiment when we plan to use an Empirical Bayes estimator, $\bskap_{2}$, to shrink the stratum causal estimates toward estimates previously obtained from an observational study. As we have shown, the risk of $\bskap_{2}$ can be computed explicitly if the stratum potential outcome variances and the stratum-specific errors of the observational study estimates,  $\bsxi$, are known. In the absence of such information, we have proposed three heuristics -- Neyman allocation, ``na\"ive" allocation assuming $\bsxi = 0$, and robust allocation under the worst-case value of $\bsxi$ given a sensitivity model -- for designing the experiment. We have also emphasized ``detachability," the ability to do good causal inference using the RCT on its own, and suggested imposing constraints on the experimental designs to ensure detachability is preserved. 

We simulated a realistic scenario in which we have access to a large observational database and are interested in a relatively rare outcome. We considered a stratification yielding twelve strata, in which the outcome frequency varies dramatically across the strata. In this setting, there were significant risk improvements to be realized by using a shrinkage estimator. Additional gains were possible when switching from an equally allocated RCT to one designed explicitly for use with a shrinkage estimator. In simulations with and without unmeasured confounding in the observational study, we saw that the na\"ive allocation typically performed well, exhibiting surprising robustness to the case when $\bsxi$ was far from zero. Our robust allocations outperformed in the case when unmeasured confounding was present in the data. 

There are many plausible extensions to this line of work. We have assumed a rigid model for treatment effect heterogeneity in this paper: treatment effects vary according to a known stratification on observed covariates. More modern work \citep{lee2021discovering, nie2021quasi} focuses on estimating heterogeneous treatment effects empirically, allowing for greater flexibility in how the effects differ across units. Incorporating such ideas into this work, we might imagine using some of the observational data in a first stage to estimate a stratification scheme, and using the remaining data for shrinkage in a second stage. Alternatively, we could consider modeling the causal effect explicitly as a function of the covariates, e.g. defining $\tau(X_i)$ rather than a vector of true treatment effects $\bstau$. We could then define a flexible shrinker to trade off between estimates not within strata, but within nearby values of the covariates themselves. 

\bibliographystyle{apalike}
\bibliography{rctodb}

\appendix 

\section{Alternatives Shrinkers to $\bskap_2$}

We have supposed that, after the experiment's conclusion, we will use  $\bskap_2$ or $\bskap_{2+}$ to shrink causal estimates from $\taur$ toward $\tauo$. However, our results are not dependent on the form of the shrinkage estimator. Our goal is to apply Theorem 3.1 in \cite{strawderman2003minimax} to compute the exact risk of the estimator. Hence, we can use any estimator $\boldsymbol \theta$ of the form 
\[ \boldsymbol \theta = \taur + \bsSig_r g_{\theta}(\taur, \tauo)\] 
where $g_{\theta}(x, y)$ is weakly differential in $x$ and $\e_r\left( || g_{\theta} ||^2 \right) < \infty$. 

A plausible alternative estimator proposed in  \cite{rosenman2020combining} is 
\[ \bskap_{1} = \taur + \left( \frac{\Tr(\bsSig_r)}{(\tauo - \taur)^\tran  \bsSig_r (\tauo - \taur)} \right) \left( \tauo - \taur \right) \,,\] 
and its positive part analogue, $\bskap_{1+}$. Two other standard alternatives, introduced in  \cite{green2005improved}, are 
\[ \bsdelt_1 = \taur + \left(  \frac{K - 2}{(\taur - \taur)^\tran  \bsSig_r^{-1} \left( \taur - \tauo \right)}\right) \left(\tauo - \taur \right) \] 
and 
\[ \bsdelt_2 = \taur + \left(\frac{(K - 2) \bsSig_r^{-1}}{(\taur - \tauo)^\tran  \bsSig_r^{-2} (\taur - \tauo)} \right)(\tauo - \taur )\,. \] 
Each also has a positive part version that is straightforward to define. 

Once a suitable estimator $\boldsymbol \theta$ is chosen, we can redefine our procedures to work with that estimator. The first step is to apply Theorem 3.1 from \cite{strawderman2003minimax} to compute its risk conditional on $\tauo$, 
\[ \mathcal{R}(\theta) = \frac{1}{K} \left( \Tr(\bsSig_r) + \e_r  \left(\sum_k \sigma_{rk}^4 \left( g_{\theta, k}^2(\taur, \tauo) + 2 \frac{\partial g_{\theta, k}(\taur, \tauo)}{\partial \tau_{rk}} \right) \right) \right)\,. \] 

Next, we can deploy the method from Section \ref{sec:exactRisk} to obtain an exact integral expression for the the risk of the estimator. \cite{bao2013moments} contains a detailed explanation of how to construct each component of the integral. For example, if we use $\bskap_{1+}$, we obtain 

\[ \mathcal{R}(\bskap_{1+}) = \frac{1}{K} \left( \Tr(\bsSig_r) + \Tr(\bsSig_r) \E\left(  \frac{4\cdot \bsnu^{\tran}\bsSig_r^{^2} \bsnu}{(\bsnu^{\tran}\bsSig_r^{} \bsnu)^2} - \frac{ \Tr(\bsSig_r) }{\bsnu^{\tran}\bsSig_r \bsnu}  \right) \right)\,, \] 
and the relevant integrals can be computed as 
\begin{equation}\label{eq:integrals}
\begin{aligned}
\E\left(  \frac{ \bsnu^{\tran}\bsSig_r^{^2} \bsnu}{(\bsnu^{\tran}\bsSig_r \bsnu)^2} \right) &= \int_0^{\infty} \det ( \boldsymbol I_K + 2 t \bsSig )^{-1/2} \cdot \exp\left( \frac{1}{2} \left(  \bsxi^\tran ( \boldsymbol I_K + 2 t \bsSig_r )^{-1} \bsxi - \bsxi^{\tran} \bsxi \right) \right) \cdot \\
& \hspace{11mm}\left( \Tr (\boldsymbol R) + (\boldsymbol L \bsSig_r^{-1/2} \bsxi)^\tran \boldsymbol R (\boldsymbol L \bsSig_r^{-1/2} \bsxi) \right) t dt \\
\E\left(  \frac{1}{(\bsnu^{\tran}\bsSig_r \bsnu)} \right) &= \int_0^{\infty} \det ( \boldsymbol I_K + 2 t \bsSig_r )^{-1/2} \cdot \exp\left( \frac{1}{2} \left(  \bsxi^\tran ( \boldsymbol I_K + 2 t \bsSig_r )^{-1} \bsxi - \bsxi^{\tran} \bsxi \right) \right) dt 
\end{aligned}
\end{equation}
where 
\begin{align*}
\boldsymbol  L &= (\boldsymbol I_K + 2 t \bsSig_r)^{-1/2}, \hspace{5mm} \text{ and } \\
\boldsymbol R &=  \boldsymbol  L ^{\tran} \bsSig_r^2 \boldsymbol{L}. 
\end{align*}

This process can be repeated for any estimator $\boldsymbol \theta$. Once integral expressions for the risk have been obtained, the design heuristics described in Section \ref{sec:designOptions} can  be deployed, using $\mathcal{R}(\boldsymbol \theta)$ as the objective function rather than $\mathcal{R}(\bskap_{2})$.

\end{document}